\documentclass[fleqn,usenatbib]{mnras}

\usepackage{graphicx}
\usepackage{amsmath}
\usepackage{amssymb}
\usepackage{newtxtext,newtxmath}

\usepackage{amsmath,amstext}
\usepackage{comment}
\usepackage[figure,figure*]{hypcap}
\usepackage{multirow}
\usepackage{amsmath}
\usepackage{nccmath}
\usepackage{rotating}
\usepackage{soul}
\usepackage{booktabs}
\setstcolor{red}
\usepackage{float}
\usepackage{longtable}
\usepackage{dcolumn}
\usepackage{orcidlink}
\usepackage{xspace}
\newcommand{\nodata}{...}

\newcommand{\rahms}[4]{$#1^{\rm h}#2^{\rm m}#3\mbox{$^{\rm s}\mskip-7.6mu.\,$}#4$} %% \dechms{04}{14}{12}{9198} = RA en formato 04^h 14^m 12^s .9198
\newcommand{\decdms}[4]{$#1^{\circ}#2'#3\mbox{$''\mskip-7.6mu.\,$}#4$} %% \decdms{28}{12}{12}{199} = Dec en formato 28^o 12' 12'' .199

\defcitealias{Ortiz2017}{OLK17}
\defcitealias{loinard2008}{LTM08}

\title[Oph-S1 near Periastron]{\centering VLBA Detections in the Oph-S1 Binary System near Periastron \\ Confirmation of its Orbital Elements and Mass}

\author[J. Ordóñez-Toro et al.]{%
\parbox{\textwidth}{
Jazmín Ordóñez-Toro$^{1}$\thanks{E-mail: n.ordonez@irya.unam.mx}\orcidlink{0000-0001-7776-498X}, 
Sergio A. Dzib$^{2}$\orcidlink{0000-0001-6010-6200}, 
Laurent Loinard$^{1,3,4}$\orcidlink{0000-0002-5635-3345}, 
Gisela Ortiz-León$^{5}$\orcidlink{0000-0002-2863-676X},
Marina A. Kounkel$^{6}$\orcidlink{0000-0002-5365-1267},
Phillip A. B. Galli$^{7}$\orcidlink{0000-0003-2271-9297}, 
Josep M. Masqué$^{8}$\orcidlink{0000-0002-1963-6848}, 
Trent J. Dupuy$^{9}$\orcidlink{0000-0001-9823-1445},
Luis H. Quiroga-Nu\~nez$^{9}$\orcidlink{0000-0002-9390-955X},
Luis F. Rodr\'{\i}guez$^{1}$\orcidlink{0000-0003-2737-5681},\\ 
}\\
$^{1}$Instituto de Radioastronom\'{\i}a y Astrof\'{\i}sica, Universidad Nacional Aut\'onoma de M\'exico, Apartado Postal 72-3, Morelia 58089, M\'exico\\
$^{2}$Max Planck Institut f\"ur Radioastronomie, Auf dem Hügel 69, D-53121 Bonn, Germany\\
$^{3}$Black Hole Initiative at Harvard University, 20 Garden Street, Cambridge, MA 02138, USA\\
$^{4}$David Rockefeller Center for Latin American Studies, Harvard University, 1730 Cambridge Street, Cambridge, MA 02138, USA\\
$^{5}$Instituto Nacional de Astrofísica, Óptica y Electrónica, Apartado Postal 51 y 216, 72000 Puebla, México\\
$^{6}$Department of Physics and Astronomy, University of North Florida, 1 UNF Dr., Jacksonville, FL, 32224\\
$^{7}$Instituto de Astronomia, Geofísica e Ciências Atmosféricas, Universidade de São Paulo, Rua do Matão, 1226, Cidade Universitária, 05508-090, São Paulo-SP, Brazil.\\
$^{8}$Departamento de Astronom\'{\i}a, Universidad de Guanajuato, Apartado Postal 144, 36000 Guanajuato, M\'exico\\
$^{9}$Institute for Astronomy, University of Edinburgh, Royal Observatory, Blackford Hill, Edinburgh, EH9 3HJ, UK\\
$^{9}$Department of Aerospace, Physics and Space Sciences, Florida Institute of Technology, 150 W University Blvd, Melbourne, 32901, FL, USA\\
}
\date{}
\usepackage{orcidlink}

\begin{document}
%\label{firstpage}
\maketitle
\begin{abstract}
Oph-S1 is the most luminous and massive stellar member of the nearby Ophiuchus star-forming region. Previous Very Long Baseline Array (VLBA) observations have shown it to be an intermediate-mass binary system ($\sim 5\,{\rm M}_\odot$) with an orbital period of about 21 months, but a paucity of radio detections of the secondary near periastron could potentially have affected the determination of its orbital parameters. Here, we present nine new VLBA observations of Oph-S1 focused on its periastron passage in early 2024. We detect the primary in all observations and the secondary at five epochs, including three within about a month of periastron passage. The updated orbit, determined by combining our new data with 35 previous observations, agrees well with previous calculations and yields masses of $4.115 \pm0.039 \,{\rm M}_\odot$ and $0.814\pm0.006 \,{\rm M}_\odot$ for the two stars in the system.
\end{abstract}


%%%%%%%%%%%%%%%%%%%%%%%%%%%%%%%%%%%%%%%%%%%%%%%%%%

%%%%%%%%%%%%%%%%% BODY OF PAPER %%%%%%%%%%%%%%%%%%

\section{Introduction}

\begin{figure*}%[!hbt]
\includegraphics[width=0.9\textwidth]{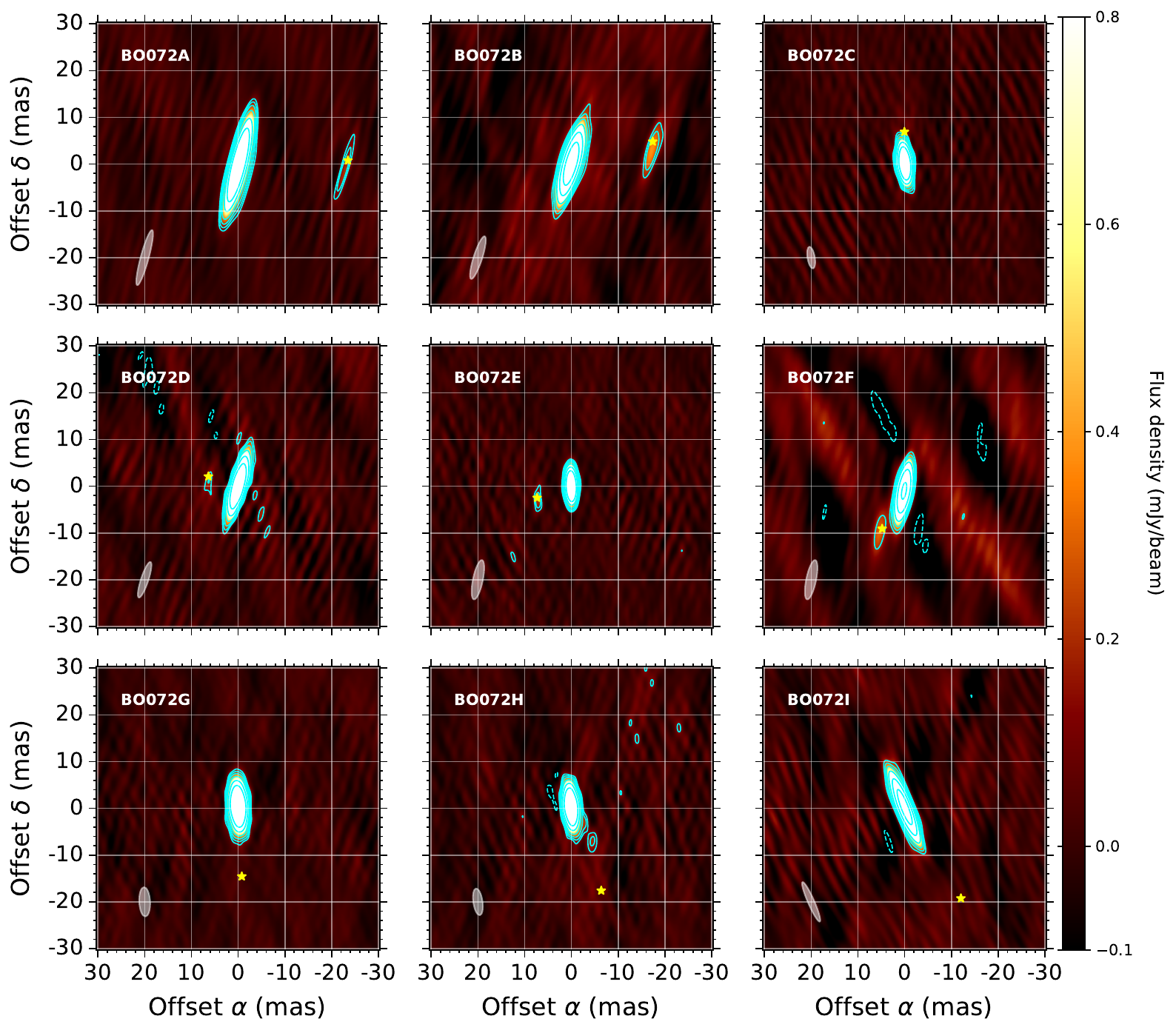}
\label{S1im}
\caption{Final radio images of S1 at 4.9 GHz corresponding to each epoch observed during the BO072 project. Intensity background images are clipped to intensities between --0.1 to 0.8 mJy beam$^{-1}$. Contour levels are at --4, 4, 6, 9, 15, 30, and 60 times the noise levels of the images, listed in Table~\ref{Ts1_full}. Images are centered on the measured position of S1A as listed in Table~\ref{Ts1_full2} per epoch, labeled in the top-left corner of each plot. The white ellipse at the bottom-left corner of each plot represents the synthesized beam of the corresponding observed epoch. Yellow stars indicate the expected position of S1B according to the astrometric results from \citet{Ordonez-Toro2024}.}
\end{figure*}

The stellar system Oph-S1 (S1 hereafter) stands out as the brightest and most massive member of the Ophiuchus star-forming region. S1 is located at a distance of 137.2 $\pm$ 0.4 pc \citep{Ordonez-Toro2024}, and it was the first young stellar object directly detected using very long baseline interferometry (VLBI) techniques \citep{andre1991}. Lunar occultation experiments in the infrared \citep{richichi94} and VLBA radio observations \citep{ortiz2017a} revealed that S1 is a binary stellar system with an angular separation on the order of 20 mas. In previous works \citep{ortiz2017a,Ordonez-Toro2024}, we took advantage of the binarity and VLBI detectability of S1 to constrain the dynamical masses of its member components \citep{Ordonez-Toro2024}. To achieve this, we combined archival Very Long Baseline Array (VLBA) observations with recent observations from our {\it Dynamical Masses of Young Stellar Multiple Systems with the VLBA} (DYNAMO--VLBA)\footnote{\url{https://www3.mpifr-bonn.mpg.de/div/radiobinaries/intro.php}} project. This resulted in a total of 35 VLBA observations distributed over 14.4 years, and enabled us to determine that the primary component, S1A, has a mass of 4.11 $\pm$ 0.10 M$_\odot$. This figure is significantly lower than the values (5 to 6 M$_\odot$) previously estimated from photometric measurements ; \citep[e.g.,][]{lada1984}. The secondary component, S1B, was found to have a mass of 0.831\,$\pm$ 0.014\,M$_\odot$, consistent with a low-mass T Tauri star.

A interesting aspect uncovered by our VLBA observations was the variability of the radio flux as a function of the orbital phase of the sources in S1 \citep{Ordonez-Toro2024}. Our analysis revealed that, while the radio emission from S1A remains relatively constant throughout the orbit, the flux from the low-mass companion S1B significantly increases near apoastron. Quantitatively, S1B was detected in 64\% of the observations carried out in the orbital phase range $0.4 \leq \phi \leq 0.6$ (apoastron corresponds to $\phi = 0.5$) while it was only detected in 24\% of the observations collected outside of the range. Noticeably, no detection of S1B has been obtained within an orbital phase 0.15 of periastron (i.e., at $\phi \leq 0.15$ or $\phi \geq 0.85$; periastron corresponds to $\phi = 0$ or $\phi = 1$). This provides an interesting puzzle since other systems (e.g.\ V773\,Tau; \citealt{torres2012}) present the exact opposite behavior (i.e., they are more radio-bright near periastron). In addition, the scarcity of detections near periastron could negatively affect the determination of the orbital elements of the system since an important portion of the orbit (periastron passages) remains comparatively poorly constrained. To examine this possibility, we conducted nine new VLBA observations of S1 focused on a periastron passage of the system. As we will see momentarily, these observations were devised to achieve significantly higher sensitivity than previous VLBA sessions and increase the probability of detections near periastron.

\section{Observations and data reduction \label{sec:obs}}
%\subsection{Observations and data reduction}

Nine radio continuum observations of S1 were conducted as part of the BO072 project using the VLBA between September 12, 2023, and June 15, 2024. Details of the observations, including the Julian date corresponding to the midpoint of each four-hour observation run, are listed in Table~\ref{Ts1_full}, while the measured positions are provided in Table~\ref{Ts1_full2}. All observations used C-band receivers at a wavelength of ($\lambda=6.0$~cm; $\nu=5\,$GHz). This wavelength was selected to balance the angular resolution, radio flux strength, and high sensitivity, which is critical to the astrometric objectives of the project.

The observational strategy followed the methodologies established in the \textit{{Gould's Belt Distances Survey }} \citep[GOBELINS; ][]{ortiz2017a} (P.I.: L. Loinard) and the DYNAMO–VLBA project (P.I.: S. Dzib). Each observation session consisted of two-minute scans of S1, bracketed by one-minute scans of the phase calibrator GBS-VLA J162700.00-242640.3 \citep{dzib2013}, also known as J1627-2427, located approximately $10'$ from S1. To reduce the impact of errors on the troposphere models used by the correlator, geodetic blocks were observed at the start, middle, and end of each session, each lasting 20 minutes.

The target was observed for a total of 120 minutes on-source, supplemented by 60 minutes for phase calibration and an additional 60 minutes for geodetic observations, resulting in a total duration of 240 minutes (4 hours) per epoch, using a recording bit rate of 4\,Gbps. The observations were designed to improve detection capabilities and allow for more frequent detections with monthly observations. Particularly, they were scheduled to cover the orbital phases from $\phi = 0.8$ to $\phi = 0.2$, passing through the periastron %passage 
($\phi=0.0$).  %With a recording bit rate of 4~Gbits~s$^{-1}$, we expected a sensitivity of 13~$\mu$Jy~beam$^{-1}$ during the 120 minutes of on-source time. This sensitivity was expected to significantly enhance detection capabilities, as previous observations of S1B indicated fluxes between 0.2 and 0.3 mJy, corresponding to levels below 10$\sigma$ . With this sensitivity, a source with a flux greater than 80~$\mu$J and at $\geq6\sigma$ can be detected.
The total observation time was approximately 4 hours $\times$ 9 epochs $=$ 36 hours.

Data calibration was carried out following the standard procedures for phase-referenced VLBI observations using the Astronomical Image Processing System (AIPS) software \citep{Greisen2003}. This process involved phase and amplitude calibration, along with corrections for group delays \citep{reid2004}, clock delays, and tropospheric effects, following the methods outlined in previous work \citep[e.g.,][]{loinard2007,dzib2010,ortiz2017a}. After the initial calibration, we applied self-calibration to further improve image quality, which allowed us to successfully detect the secondary component with greater clarity; the detailed steps of this process are described in \citet{Ordonez-Toro2024}.

Following calibration, the visibility data were imaged using natural weighting (ROBUST = 5 in AIPS). The resulting rms noise levels in the final images ranged between (25--52~$\,\mu$Jy beam$^{-1}$, see Table~\ref{Ts1_full}). The positions and flux densities of the detected sources were measured using a two-dimensional Gaussian fitting procedure (JMFIT task in AIPS). Table~\ref{Ts1_full2} presents the complete list of measured positions, and Figure \ref{S1im} shows all the images of S1 from the BO072 project. 

\begin{table}%[ht]
    \caption{Observing logs for the BO072 project.}
    \label{Ts1_full}
    \footnotesize
    \begin{center}
    \setlength{\tabcolsep}{2.6pt}
    \renewcommand{\arraystretch}{1.05}
    \begin{tabular}{lcccc} \hline\hline
    Ep. & Date UT & Julian Date & \multicolumn{1}{c}{Synthesized Beam} & $\sigma_{\rm noise}$  \\  
    & & & $\theta_{\text{maj}} \times \theta_{\text{min}}$; P.A. &  \\  
    & (yyyy.mm.dd) & & (mas $\times$ mas; deg) & ($\mu$Jy\,bm$^{-1}$)  \\ \hline
    A & 2023.09.12 & 2460199.5313 & 12.25$\times$1.72; -15.15 & 25 \\
    B & 2023.10.24 & 2460242.3757 & 9.59$\times$1.84; -17.68 & 43  \\
    C & 2023.12.30 & 2460309.2132 & 4.73$\times$1.57; 9.53 & 33    \\
    D & 2024.01.21 & 2460331.1535 & 8.03$\times$1.60; -18.13 & 35  \\
    E & 2024.02.04 & 2460345.1146 & 8.61$\times$2.12; -11.63 & 27  \\
    F & 2024.02.29 & 2460370.0743 & 8.61$\times$2.12; -11.63 & 52  \\
    G & 2024.04.04 & 2460404.9514 & 6.30$\times$2.33; 2.73 & 31    \\
    H & 2024.05.08 & 2460438.8361 & 5.79$\times$1.98; 8.15 & 30    \\
    I & 2024.06.15 & 2460476.7542 & 9.24$\times$1.35; 23.66 & 48   \\
    \hline
    \end{tabular}  
    \end{center}
\end{table}

\begin{figure}%[!h]
\includegraphics[scale=0.32]{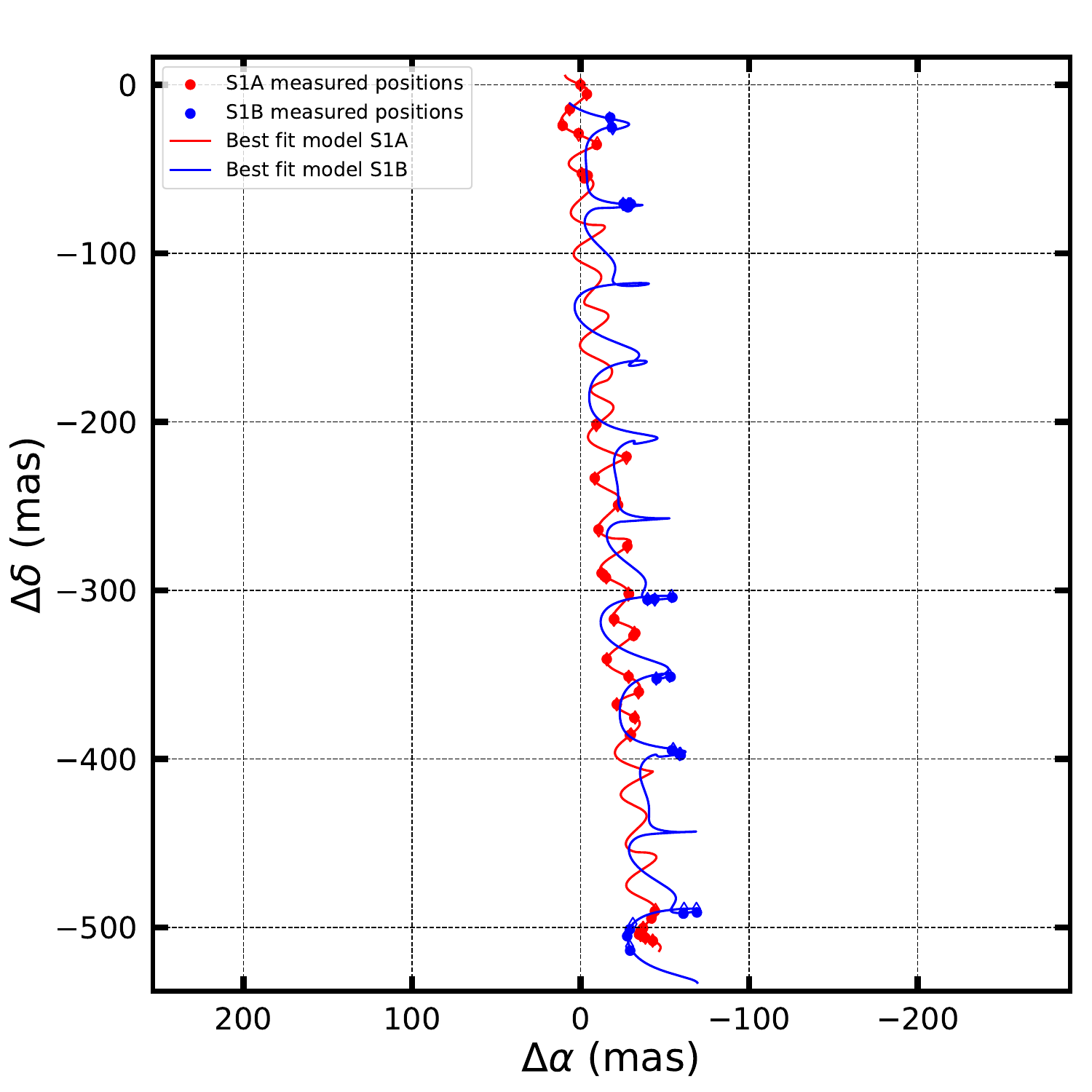}\\
\includegraphics[scale=0.32]{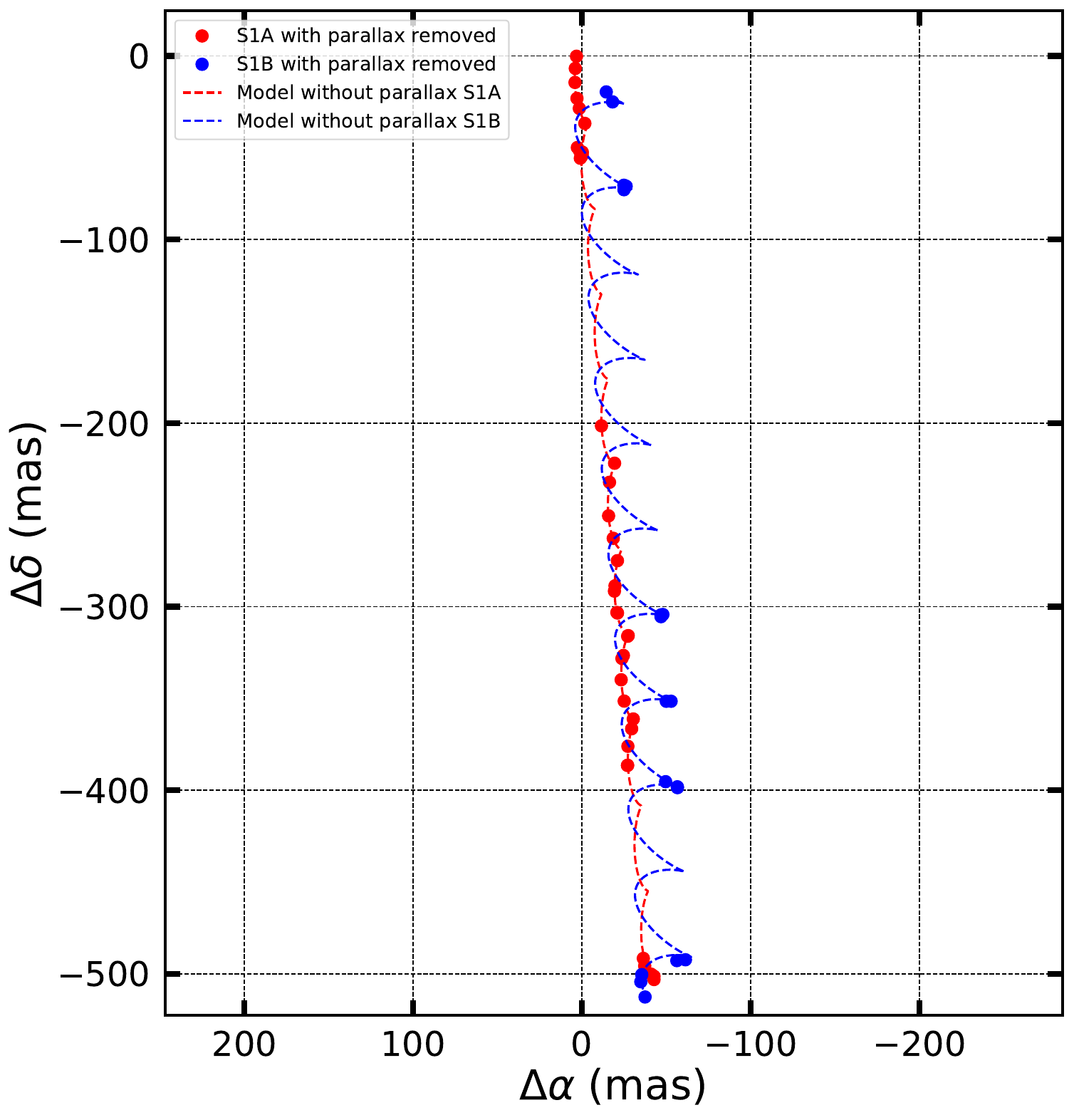}\\
\label{S1sky}
\caption{Measured positions of S1A (red dots) and S1B (blue dots) shown as offsets from the position of S1A on June 24, 2005 ($\alpha=$\rahms{16}{26}{34}{17392792}; $\delta$=\decdms{-24}{23}{28}{428288}). 
Top panel: The red and blue curves show the stellar motion best fits, described in the text, to the positions of S1A and S1B, respectively.  Bottom panel: The measured positions and best fit after removing the parallax signature.}
\end{figure}

\section{Results}

\begin{table*}%[H!]
    \caption{S1A and S1B measured positions and flux densities from the VLBA observations of the BO072 project.}
    \label{Ts1_full2}
    \footnotesize
    \begin{center}
    \setlength{\tabcolsep}{2.6pt}
    \renewcommand{\arraystretch}{1.05}
    \begin{tabular}{lcccccccc} \hline\hline
     & \multicolumn{3}{c}{S1A} & \multicolumn{3}{c}{S1B} & \multicolumn{1}{c}{} \\ 
    \cmidrule(lr{.75em}){2-4} \cmidrule(lr{.75em}){5-7}
    Ep. & $\alpha$(J2000.0) & $\delta$(J2000.0) & $S_\nu$ & $\alpha$(J2000.0) & $\delta$(J2000.0) & $S_\nu$ $^{\mathrm{a}}$ & $\phi^{\mathrm{b}}$ \\  
    & $16^{h}26^{m}$ [$^{s}$] & $-24^{\circ}23'$ [$''$] & (mJy) & $16^{h}26^{m}$ [$^{s}$] & $-24^{\circ}23'$ [$''$] & (mJy) & \\ \hline
    A & 34.17098493(59) & 28.918702(27) & $~~6.10\pm0.05$ & 34.1693223(145) & 28.919447(669) & $0.22\pm0.04$ & 0.795 \\
    B & 34.17112513(100) & 28.923003(37) & $~~6.82\pm0.09$ & 34.1698558(122) & 28.920134(453) & $0.54\pm0.09$ & 0.862 \\
    C & 34.17145174(26) & 28.928744(10) & $~~8.66\pm0.06$ & \nodata & \nodata & $<0.10$ & 0.968 \\
    D & 34.17151614(55) & 28.930225(21) & $~~7.58\pm0.08$ & 34.1719746(100) & 28.929418(398) & $0.26\pm0.06$ & 0.002 \\
    E & 34.17156010(22) & 28.930536(9) & $~~8.50\pm0.06$ & 34.1720807(49) & 28.933517(225) & $0.32\pm0.05$ & 0.024 \\
    F & 34.17160976(107) & 28.932568(43) & $~~5.99\pm0.12$ & 34.1719630(113) & 28.942077(459) & $~~1.05\pm0.15$ & 0.064 \\
    G & 34.17155446(34) & 28.932828(13) & $~~7.11\pm0.05$ & \nodata & \nodata & $<0.09$ & 0.119 \\
    H & 34.17136367(34) & 28.934420(13) & $~~6.65\pm0.05$ & \nodata & \nodata & $<0.09$ & 0.172 \\
    I & 34.17106563(117) & 28.936320(37) & $~~7.13\pm0.10$ & \nodata & \nodata & $<0.14$ & 0.232 \\
    \hline
    \end{tabular}  
    \end{center}
    {\raggedright \footnotesize
    Note: The number in parentheses in the right ascension and declination columns indicate the statistical uncertainties on the final digits of the corresponding values. To obtain the complete uncertainty 0.24 mas and and 0.46 mas must be added quadratically in right and declination, respectively (see text).\\
    $^{\mathrm{a}}$ Upper limits for S1B correspond to three times the image noise level. $^{\mathrm{b}}$ $\phi$ is the orbital phase at each epoch.
    \par}
\end{table*}

The new observations from the BO072 project (Table~\ref{Ts1_full2}), which focused on detecting the component S1B near the periastron,
enabled us to successfully identify S1B in five out of the nine observation epochs, as shown in Figure~\ref{S1im}.
By using the astrometric results from \citet{Ordonez-Toro2024}, we estimated the expected position of S1B in all
nine observations and, as can be seen in Fig.~\ref{S1im}, the positions of S1B in our new detections are in good agreement with the expectations. 
The radio source associated with S1A was detected in all observed epochs.
The mean fluxes are 7.17 and 0.48 mJy for S1A and S1B, respectively. These results are consistent
with those obtained from previous VLBA observation campaigns of S1 \citep[e.g.][]{ortiz2017a,Ordonez-Toro2024}.
These new observations, combined with previous VLBA data from the GOBELINS and DYNAMO-VLBA projects (see Table 1 of \citealt{Ordonez-Toro2024}), enabled a comprehensive analysis with a total of 44 observational epochs, from June 2005 to June 2024 (almost 19 years). 
We performed an astrometric fit based on all the measured positions of S1A and S1B.

The motion of binary stars across the sky is characterized by their trigonometric parallax $\pi$, the uniform proper motion of their center of mass in right ascension ($\mu_\alpha$) and declination ($\mu_\delta$), along with the contributions from their orbital dynamics. The equations for the primary star are expressed as follows:

\begin{eqnarray}
\alpha(t) & = & \alpha_0 + \mu_\alpha t + \pi f_\alpha(t) + a_1 Q_\alpha(t),\label{eqn:pm1}\\
\delta(t) & = & \delta_0 + \mu_\delta t + \pi f_\delta(t) + a_1 Q_\delta(t)\label{eqn:pm2}.
\end{eqnarray}

\noindent
In these equations, $f_\alpha(t)$ and $f_\delta(t)$ correspond to the projections of the parallax ellipse in right ascension and declination, while $Q_\alpha(t)$ and $Q_\delta(t)$ represent the projections of the orbital motions. The detailed expressions are provided in \citet{Ordonez-Toro2024}.

\begin{figure}
\label{S1or}
\includegraphics[scale=0.25]{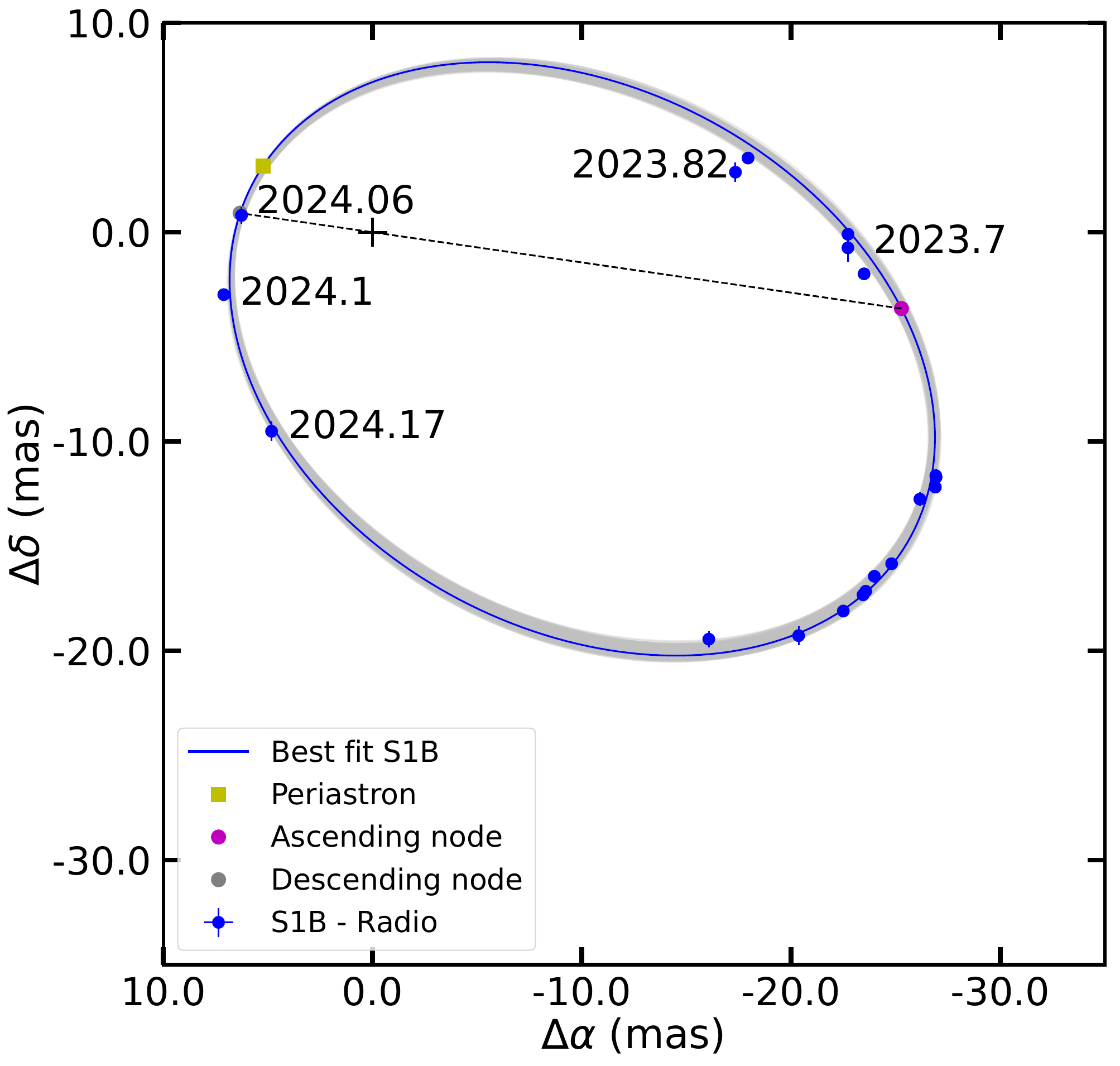}
\caption{Orbit of the S1 binary system shown in terms of the relative positions of S1B (blue points) with respect to S1A. The error bars account for the uncertainties in the positions of both stars, calculated by combining their individual errors in quadrature. The dashed black line traces the predicted line of nodes from the model, and the black cross marks the location of the primary component, S1A.}
\end{figure}

To estimate the astrometric and orbital parameters of the S1 system, we follow the same procedure as in \citet{Ordonez-Toro2024} and applied the MPFIT least squares fitting algorithm \citep[see also][and references therein]{kounkel2017}. This approach fits the observed positions to the equations of motion and allows us to determine the individual masses of the stars by accounting for their motion relative to the center of mass.
%To complement these results, we also employed the \orbitize package, which combines the \textit{Orbits For The Impatient} (OFTI) methods with a \textit{Markov Chain Monte Carlo} (MCMC) fitting scheme \citep{orbitize}. While this method primarily focuses on relative positions and requires the system distance as an input parameter, we fixed this distance based on the MPFIT results. Although \orbitize provides only the total mass of the system, it offers a rigorous treatment of uncertainties, particularly suitable for VLBI data sets, as demonstrated in previous work \citep{Ordonez-Toro2024}.
The results of the best fit are summarized in Table~\ref{T_A0}, while the visualizations of the measured positions and the best fits are shown in Figure~\ref{S1sky}. For easier visualization, that figure shows both the complete motion (top panel), and the motion with the parallax component removed.
The orbit solution and relative positions of S1B with respect to S1A are presented in Figure~\ref{S1or}. The fitting analysis yields a mass of $4.115  \pm 0.039 $\,M$_\odot$ for the primary component S1A and $0.814 \pm 0.006 $\,M$_\odot$ for the secondary component S1B, resulting in a total system mass of $4.911 \pm 0.048$ M$_\odot$. It is important to note that a degeneracy exists between the parameters $\omega$ and $\Omega$ due to the lack of radial velocity measurements. This arises from the invariance of the astrometric equations under such the transformation $\omega \rightarrow \omega+180^\circ$, $~\Omega \rightarrow \Omega+180^\circ$. 

The statistical errors on positions listed in Table~\ref{Ts1_full2} are those delivered by the image fitting procedure and only account for the noise level in the images \citep{condon1997}. Residual phase errors in the calibrated visibilities, however, introduce additional (systematic) astrometric uncertainties which typically dominate the total error budget \citep[e.g.][]{loinard2007}. These systematic errors depend, among other things, on the typical elevation during the observations. In Ophiuchus, which barely reaches elevations above 20 degrees at most VLBA sites, they can be fairly large. To quantify them, we iteratively searched for the amounts to be quadratically added to the statistical errors listed in Table~\ref{Ts1_full2} until a reduced $\chi^2$ value of one is reached (separately in right ascension and declination). This is a standard practice previously employed, for instance, in \citet{loinard2007} and \citet{Menten2007}. In the case of S1, we need to treat separately the 14 first observations (obtained between 2005 and 2007) from the more recent ones, because the former used a different gain calibrator (the systematic errors depend strongly on the separation between the target and calibrator -- \citealt{reid2017}). We find that we need to add systematic errors of 0.85 mas and 1.25 mas, in right ascension and declination, respectively, for the first 14 epochs. For the more recent observations, the corresponding figures are 0.24 mas and 0.46 mas. For completeness, we note that the uncertainties reported on the fitted parameters do account for the systematic contributions to the errors.

%For the analysis with \orbitize, we utilized 19 observations in which both components, S1A and S1B, were detected simultaneously. The position offsets and position angles of the components served as inputs for \orbitize, with prior definitions established for the orbital elements. We conducted an MCMC exploration with 10,000 walkers over 10,000 orbital configurations. The results are summarized in Table~\ref{mcmc}, and a corner plot displaying the posterior distributions is included in Figure~\ref{corner_plot} of the appendix \ref{app:1}. Additionally, Figure~\ref{plot_mcmc} shows 2,000 orbits of the allowed orbital configurations for S1B, derived from the MCMC analysis. The resulting total mass of the binary system is $5.144^{+0.068}_{-0.048}$ M$\odot$.

\begin{table}
	\renewcommand{\arraystretch}{1.1}
%	\begin{minipage}{0,7\textwidth}
		\caption{Best-fit model parameters for the close binary system S1.}		\label{T_A0}
%	\end{minipage}
	\centering
	%\footnotesize
	\small
	\begin{tabular}{cccc} \hline \hline
		&Parameter& Value & Units 	  \\ 
		\hline
		\multicolumn{2}{l}{Astrometric parameters} &&\\
		& $\alpha_{2016.0, {\rm centre}}$ & 16:26:34.1723733(8)& hh:mm:ss\\
		%& $\sigma_\alpha$ & 0.0000023& ss\\
		& $\delta_{2016.0, {\rm centre}}$ & --24:23:28.7131602(85) & $^\circ:\,':\,''$\\
		%& $\sigma_\delta$ & 0.000022 & $''$\\
		%&$\Delta\alpha_{\rm S1A}$\tablenotemark{a}	& $3.69\pm0.04$&mas\\
        %&$\Delta\alpha_{\rm S1A}$& $3.69\pm0.04$&mas\\
		%&$\Delta\delta_{\rm S1A}$& $3.25\pm0.02$&mas\\
		%&$\Delta\alpha_{\rm S1B}$& $-17.99\pm0.04$&mas\\
		%&$\Delta\delta_{\rm S1B}$&$-15.87\pm0.02$ &mas\\
		&$\mu_{\alpha}$  & $-2.486\pm0.002$ & mas yr$^{-1}$\\	
		&$\mu_{\delta}$  & $-26.756\pm0.001$ & mas yr$^{-1}$\\
		&$\pi$           & $7.30\pm0.02$ & mas\\
		&d & $137.03\pm0.32$ & pc \\
		\multicolumn{2}{l}{Orbital parameters}&&\\
		&$a_1$ & $0.406\pm0.001$ & AU \\
		&$a_2$ & $2.053\pm0.007$ & AU \\
	    &P & $1.737\pm0.001$ & years\\
	    &$T_0$ & $2457162.33\pm0.45$ & Julian date \\
	    &$e$ & $0.657\pm0.002$ & \\
	    &$\Omega$ & $261.84\pm7.70$ & degrees \\
	    &$i$ & $20.0\pm2.3$ & degrees \\
	    &$\omega$ & $155.9\pm7.8$ &degrees \\
	    \multicolumn{2}{l}{Dynamical masses}&&\\
	    &Total mass & $4.929\pm0.045$ & M$_\odot$ \\
	    &Mass 1 & $4.115\pm0.039$ & M$_\odot$ \\
	    &Mass 2 & $0.814\pm0.006$ & M$_\odot$ \\
	    \hline
	\end{tabular}  
%\tablenotetext{a}{$\Delta\alpha$ and $\Delta\delta$ values are the orbital offset relative to the mass center at the epoch 2016.0 of components S1A and S1B.  }
\end{table} 

%\begin{table}
%	\renewcommand{\arraystretch}{1.1}
%\caption{Best-fit model parameters for the close binary system S1 using \orbitize}
%	\label{mcmc}
%	\centering
%	\small
%	\begin{tabular}{cccc} \hline \hline
%		& Parameter & Value & Units \\ 
%		\hline
%		\multicolumn{2}{l}{Orbital Parameters}&&\\
%        & $a$ & 2.493$^{+0.011}_{-0.008}$ & AU \\
%        & $e$ & 0.669$^{+0.003}_{-0.004}$ & \\
%        & $\Omega$ & 262.016$^{+2.611}_{-1.646}$ & degrees \\
%        & $i$ & 26.975$^{+0.979}_{-0.957}$ & degrees \\
%        & $\omega$ & 154.021$^{+1.897}_{-2.953}$ & degrees \\
%        & $\tau$ & 2458850.081$\pm{0.002}$ & \\
%        & P & 1.736$\pm{0.001}$  & years \\ 
%        & Mass (A+B) & 5.144$^{+0.068}_{-0.048}$ & $M_{\odot}$ \\
%        %& Parallax & 7.299$^{+0.003}_{-0.003}$ & mas \\
%	    \hline
%	\end{tabular}  
%\end{table}

%%\begin{figure*}%[!hbt]
%%\includegraphics[width=\textwidth]{Imagenes/Orbit_mcmc_S1.png}\\
%%\label{plot_mcmc}
%%\caption{Allowed orbital configurations for the binary system S1, derived from the MCMC analysis conducted using the Orbitize! library. The black marker (star) designates the position of the primary component, and the red points represent radio observations of the secondary component. The upper right panel depicts the evolution of the angular separation over time, while the lower right panel illustrates the variation in the position angle throughout the orbits.}
%%\end{figure*}

%\subsection{Flux variations}

%%%%%%%%%%%%%%%%%%%%%%%%%%%%%%%%%%%%%%%%%%%%%%%%%%%%%%%%%%%%%%%%%%%%%%%%%
\section{Discussion and Conclusions}

The new VLBA observations of S1 reported here have enabled the detection of the secondary component, S1B, at five epochs near the periastron of the system. This includes three detections between $0.0 \leq \phi \leq 0.1$ where no detections have been obtained before. The measured positions of S1B at these epochs coincide very well with the predictions of the orbit model previously reported by \citet{Ordonez-Toro2024} demonstrating that the paucity of detections near periastron in our previous work did not negatively affect the orbit determination. As a consequence, the astrometric and orbital parameters derived here are fully consistent with our previous determination, but with improved uncertainties. The masses obtained here are $M_A = 4.115 \pm0.039 $\,M$_\odot$ and $M_B = 0.814\pm0.006 $\,M$_\odot$ compared to $M_A = 4.11\pm0.10$\,M$_\odot$ and $M_B = 0.831\pm0.014$\,M$_\odot$ in \citet{Ordonez-Toro2024}. These results confirm a mass for S1A about 25\% lower than estimated from its photometric properties. S1B, on the other hand, is confirmed to be a Solar-mass T Tauri star.

Our observations also provide new clues on the radio behavior of S1\,B. As mentioned above, \citet{Ordonez-Toro2024} documented a lower detection rate of S1B near the periastron than close to apoastron. Since a decrease in magnetic activity on S1B near the periastron is difficult to envision and is contrary to the situation in other systems \citep[e.g.,][]{torres2012}, they considered the possibility that S1A could be surrounded by an optically thick region, where S1B would plunge when its separation from S1A becomes smaller than 20 mas. Our new observations definitely discard this possibility since we obtained detections of S1B very near periastron (our detection at epoch January 21, 2024 corresponds to $\phi\simeq0.002$). We note, finally, that the detection rate of S1B in our new data (55\%; five detections out of nine observations) is close to the value reported by \citet{Ordonez-Toro2024} near apoastron (64\%). However, this similarity should not be interpreted as evidence that no radio difference exist between periastron and apoastron since the new observations are, on average, significantly more sensitive than the previous ones.

\section*{Acknowledgements}

J.O., L.L., G.O.L. and J.M.M. acknowledges the support of DGAPA PAPIIT grants IN112416, IN108324 and IN112820 as well as CONACyT-CF grant 263356.  S.A.D. acknowledge the M2FINDERS project from the European Research Council (ERC) under the European Union's Horizon 2020 research and innovation programme (grant No 101018682). P.A.B.G. acknowledges financial support from the São Paulo Research Foundation (FAPESP) under grant 2020/12518-8.
The National Radio Astronomy Observatory is a facility of the National Science Foundation operated under cooperative agreement by Associated Universities, Inc.  This document was prepared using the collaborative tool Overleaf available at \url{https://www.overleaf.com/}. 

%\facilities{VLBA \citep{napier1994}, Gaia \citep{gaia2016}} 
%\software{AIPS \citep{Greisen2003}, Astropy\footnote{\url{https://www.astropy.org/}} \citep{astropy2013}, NumPy\footnote{\url{https://www.numpy.org/}} \citep{numpy2011}, SciPy\footnote{\url{https://www.scipy.org/}} \citep{scipy2014}, Matplotlib\footnote{\url{https://matplotlib.org/}} \citep{matplotlib2007}, and APLpy \citep{aplpy2012}}
%%%%%%%%%%%%%%%%%%%%%%%%%%%%%%%%%%%%%%%%%%%%%%%%%%%
\section*{Data Availability}

The data underlying this article will be shared on reasonable request to the corresponding author.

\label{lastpage}
\end{document}